\documentclass[fleqn,usenatbib]{mnras}

\usepackage{newtxtext,newtxmath}

\usepackage[T1]{fontenc}

\DeclareRobustCommand{\VAN}[3]{#2}
\let\VANthebibliography\thebibliography
\def\thebibliography{\DeclareRobustCommand{\VAN}[3]{##3}\VANthebibliography}

\usepackage{graphicx}	
\usepackage{amsmath}	

\newcommand{\pdv}[2]{\frac{\partial #1}{\partial #2}}

\title[Cadence effects in exoplanet transits]{Understanding and predicting cadence effects in the characterization of exoplanet transits}

\author[Camero, Ho, \& Van Eylen]{
Julio Hernandez Camero$^{1,2}$\thanks{E-mail: julio.camero.21@ucl.ac.uk}, Cynthia S. K. Ho$^{2}$,
Vincent Van Eylen$^{2}$
\\
$^{1}$Department of Physics and Astronomy, University College London, Gower Street, London WC1E 6BT, UK \\
$^{2}$Mullard Space Science Laboratory, University College London, Dorking RH5 6NT, UK
}

\date{Accepted XXX. Received YYY; in original form ZZZ}

\pubyear{2023}

\begin{document}
\label{firstpage}
\pagerange{\pageref{firstpage}--\pageref{lastpage}}
\maketitle

\begin{abstract}
We investigate the effect of observing cadence on the precision of radius ratio values obtained from transit light curves by performing uniform Markov Chain Monte Carlo fits of 46 exoplanets observed by the Transiting Exoplanet Survey Satellite (TESS) in multiple cadences. We find median improvements of almost 50\% when comparing fits to 20s and 120s cadence light curves to 1800s cadence light curves, and of 37\% when comparing 600s cadence to 1800s cadence. Such improvements in radius precision are important, for example, to precisely constrain the properties of the radius valley or to characterize exoplanet atmospheres. We also implement a numerical Information Analysis to predict the precision of parameter estimates for different observing cadences. We tested this analysis on our sample and found it reliably predicts the effect of shortening observing cadence with errors in the predicted \% precision of $\lesssim 0.5 \%$ for most cases. We apply this method to 157 TESS object of interest that have only been observed with 1800s cadence to predict the precision improvement that could be obtained by reobservations with shorter cadences and provide the full table of expected improvements.  We report the 10 planet candidates that would benefit the most from reobservations at short cadence. Our implementation of the Information Analysis for the prediction of the precision of exoplanet parameters, Prediction of Exoplanet Precisions using Information in Transit Analysis (\texttt{PEPITA}) is made publicly available.
\end{abstract}

\begin{keywords}
exoplanets -- methods: numerical -- methods: observational -- planets and satellites: general -- planets and satellites: detection
\end{keywords}



\section{Introduction}

The transit technique has proven its success with more than 3500 exoplanets discovered using it. Missions like Kepler \citep{boruckiKeplerPlanetDetectionMission2010, 2007ASPC..366..309B, boruckiKEPLERMissionDevelopmentOverview2016} and the Transiting Exoplanet Survey Satellite (TESS) \citep{rickerTransitingExoplanetSurvey2015} have provided the exoplanet community not only with a large number of newly discovered exoplanets, but also with the ability to characterize them with an ever-increasing level of detail as we better understand how to extract the information contained in their transits.

The transit of an exoplanet allows us to measure the radius ratio between the planet and the star, $R_{p}/R_{*}$ \citep[see e.g.][]{seagerUniqueSolutionPlanet2003} from which, knowing the radius of the star, one can determine the radius of the planet. There is a particular interest in obtaining precise measurements of this quantity, since knowledge about an individual exoplanet or an exoplanet population can be derived given such measurements of this quantity. For example, using data from the Kepler mission, a drop in the number of exoplanets with radii in between Earth's and Jupiter's was discovered \citep{fultonCaliforniaKeplerSurveyIIIGap2017}. This valley is known as the ``radius valley'' and was already predicted before its discovery by several groups \citep[see e.g.][]{owenKeplerPlanetsTale2013}. However, the availability of precise planetary radii measurements will be essential in the correct characterization (position and depth) of the valley \citep[see e.g.][]{Ho_2023, vaneylenAsteroseismicViewRadius2018, huber20SecondCadence2022}. Planetary radii are also essential in obtaining estimates for planetary densities, which are indicative of the composition of an exoplanet \citep[see e.g.][]{zengSimpleAnalyticalModel2017}. Moreover, extremely precise measurements of transit depths (which is directly related to planetary radii), with precisions of ~$0.5\%$ in different bands can be used to obtain transmission spectra of atmospheres and thus, infer the atmospheric composition of exoplanets \citep[see e.g.][]{yangRevisitingKELT19AbWASP156b2022}.

\begin{figure}
  \centering
  \includegraphics[width=0.95\columnwidth]{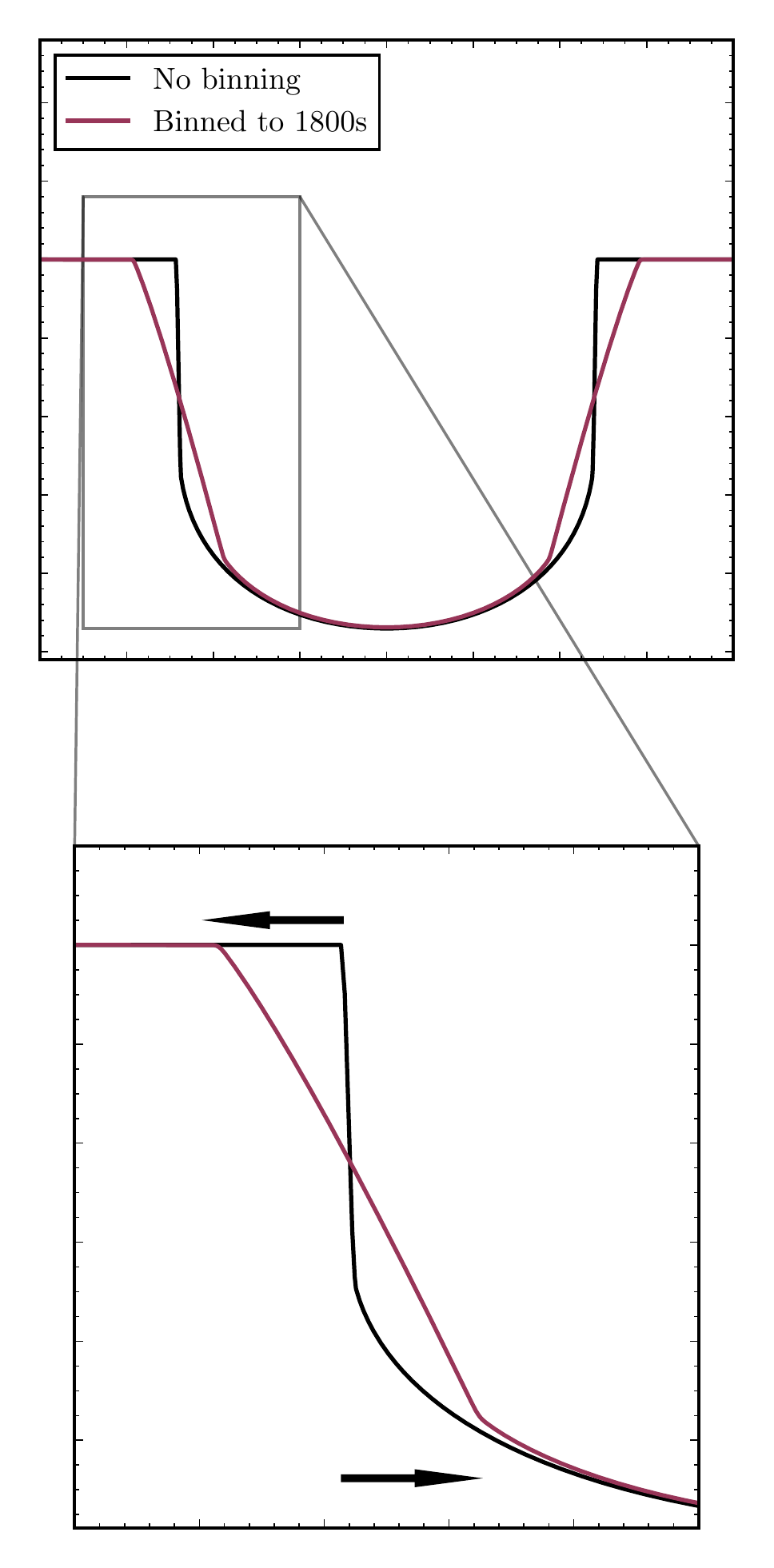}
  \caption{The binning of a light curve produces deformations in its morphology. These are most evident in the shift of the contact points and the lengthening of the ingress and egress here highlighted with arrows in the zoomed plot. These deformations are understood and can be predicted by models as evidenced by the light curves shown in this figure which were generated using \texttt{PyTransit}.}
  \label{fig:cadence.pdf} 
\end{figure}

\begin{figure*}
  \centering
  \includegraphics[width=1.0\textwidth]{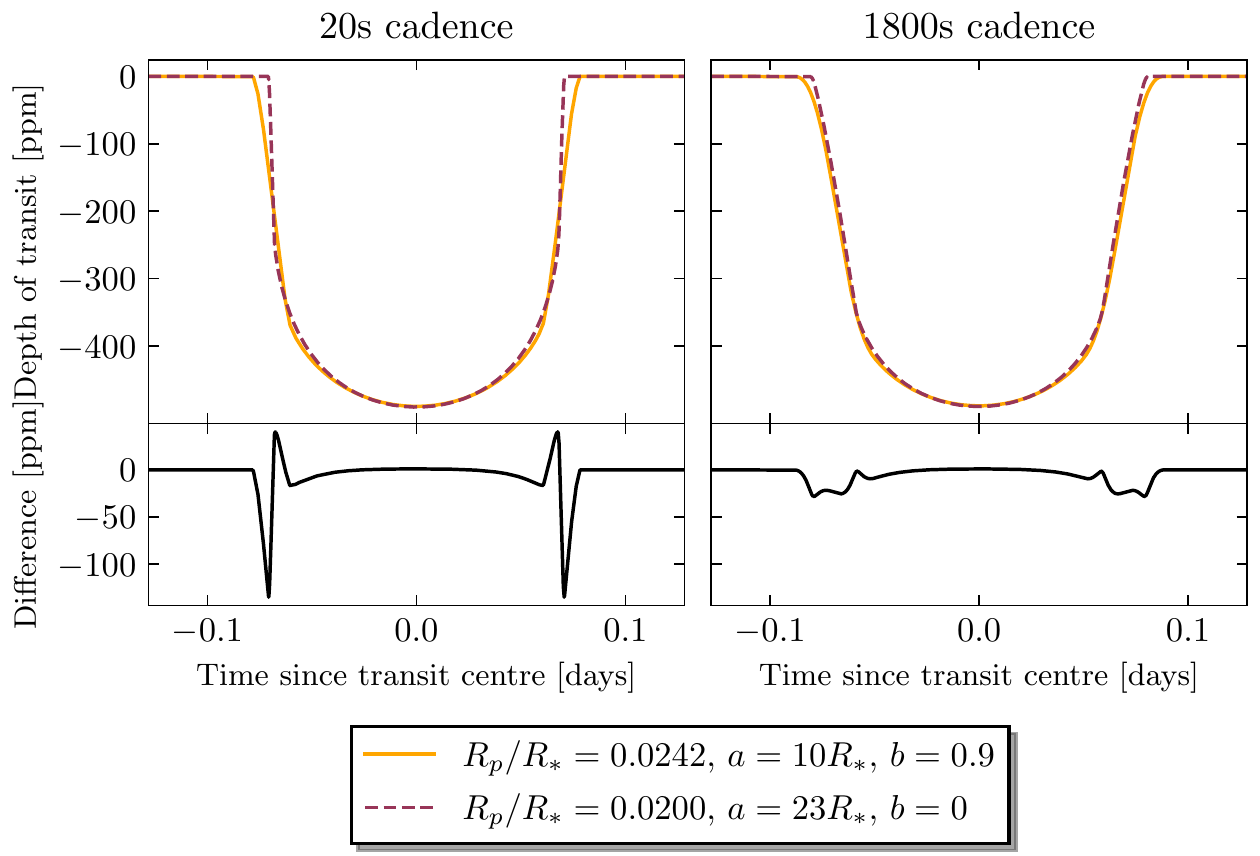}
  \caption{Transit of planets with two different sets of parameters in both short (20s) and long (1800s) cadence. The bottom plots show the differences between the light curves and highlight how the longer cadence light curves have smaller difference (are more alike) and how, in this way, information that was contained in the shorter cadence light curve has been lost. The differences in radius ratio here would correspond to an ambiguity between the transit produced by a planet with a radius of $\sim 2.18 R_\oplus$ and a planet with a radius of $\sim 2.64 R_\oplus$ for a Sun-like star, an increase of radius of $21\%$. Light curves are generated using \texttt{PyTransit} and the plot is based on the ideas presented by \protect\cite{petiguraTwoViewsRadius2020}.}
  \label{fig:two_cadences.pdf} 
\end{figure*}

In the past years, concerns have been raised about the possible influence that the choice of cadences in the observation of exoplanet transit light curves may have in the precision of parameters derived from these events \citep[see e.g.][]{dawsonPhotoeccentricEffectProtohot2012, petiguraTwoViewsRadius2020, huber20SecondCadence2022, alexoudiParameterRefinementInflated2022}. Even before this, \cite{kippingBinningSinningMorphological2010} described how the use of longer cadences introduces distortions in the morphology of a transit light curve, affecting mostly the ingress and egress of the transit as shown in Figure \ref{fig:cadence.pdf}. However, these deformations of the light curve can be modelled by numerically integrating a light curve to the required cadence \citep{kippingBinningSinningMorphological2010}. Presently, the concern lies in how the use of longer cadences (i.e. integrating the light curves to longer times) represents a loss of information that cannot be recovered even though our models accurately predict the shape of binned light curves \citep[see e.g.][]{petiguraTwoViewsRadius2020}. This loss of information translates into light curves arising from different parameter sets becoming more alike and thus, reducing the precision with which we may extract information about different parameters from these light curves. Figure \ref{fig:two_cadences.pdf} illustrates this idea with the transits of two planets with a different set of parameters, both with $20$s and $1800$s cadence. The bottom plots of the Figure show the difference between the light curves produced by each of the planets in each cadence and illustrate how, for the shorter cadence, there is a greater difference between the transits, while for the longer cadence the differences are smaller, which is what is meant by saying that information is lost. 

This idea of the information contained in a light curve has been explored in previous works such as \citet{carterAnalyticApproximationsTransit2008, priceTransitlightcurves2014}, where they implement an analytical Fisher Information Analysis (Information Analysis henceforth) to approximate, non-limbdarkened forms of transit light curves. The analysis of \citet{priceTransitlightcurves2014} allows them to predict, for each cadence and particular parameter set, what is the best precision that can be obtained by fitting each of the model parameters. 

In this work, we approach the issue of cadence from two sides. First, we aim to provide an in-depth analysis of its effects by performing fits to a number of TESS confirmed planets that were observed using more than one cadence and then compare the resulting precisions in the radius ratio to understand what the impact of cadence is. Second, we extend previous analytical implementations of the Information Analysis by developing a numerical implementation of the analysis that can be applied to the non-approximate and limb-darkened forms of transit light curves and that can be adapted to any fitting model. We compare the predictions of this analysis with the results of our previous fits to understand whether the Information Analysis may be used as a reliable tool in the prediction of parameter precisions and compare our method with previous analytical methods.

In Section \ref{sec:method} we lay out the methodology that we follow in order to perform the homogeneous fits of a large number of light curves as well as the process of candidate selection. We also present our implementation of the Information Analysis to exoplanet transits. Then, in Section \ref{sec:results} we summarize the results obtained by our work which are then discussed in Section \ref{sec:discussion}, and we conclude in Section \ref{sec:conclusion}.

\section{Method}
\label{sec:method}

\subsection{Candidate selection} 
\label{sub:candidate_selection}

The selection of candidates starts with the full list of TESS confirmed planets obtained from the NASA Exoplanet Archive (\cite{nasaPlanetarySystemsTable2020}, downloaded June 2022), henceforth NEA, out of which systems consisting of a single planet orbiting a single star are selected, reducing the original number of 231 systems down to 105. We make this choice for computational simplicity and to reduce the number of light curves that need to be fitted. We do not expect that the presence of other planets in the system and/or possibly the presence of more than one star in the system to invalidate the results here presented, although these cases would require an extension of the prediction algorithm presented below to simultaneously model multiple planets and account for third light contamination. However, this is something that should be investigated independently. For each of these systems, we obtain the available light curves using the Lightkurve python package \citep{2018ascl.soft12013L} to search for light curves in the MAST data archive\footnote{\url{https://archive.stsci.edu/index.html}}. In order to ensure the homogeneity in the treatment applied to the light curves, only those authored by the TESS Science Processing Operations Center or SPOC---which is in charge of receiving raw data and extracting photometry and astrometry for each target and identifying and removing systematic errors among other tasks---are used. For a detailed description of the TESS SPOC pipeline, see \cite{jenkinsTESSScienceProcessing2016} and the data release notes\footnote{\url{https://archive.stsci.edu/tess/tess_drn.html}}. Once a list of all the available light curves along with their cadences has been obtained for all the systems, only those for which more than a single cadence is available are selected in this step, reducing the list down to 83 systems. This is done because light curves released with a shorter cadence do not necessarily have longer cadence light curves released by SPOC.

When a system has been observed in multiple cadences, since some regions of the sky were observed in more TESS sectors than others, different cadences for a given system may have a different number of available sectors. That is, out of all cadences available so far for TESS light curves (i.e. 20s, 120s, 600s and 1800s), a target may have been observed with 1800s in five sectors, with 600s in four and with 120s in just one. Therefore, we fit sectors of a given exposure one at a time in order to allow for a more homogeneous comparison of different cadences so that any changes in the retrieved parameters can be assumed, at least to the extent of the precautions taken here allow for, to arise from cadence effects. Whether this decision may have an influence in our results is explored later on by fitting a small number of systems with the same number of sectors available for several cadences.

However, the decision to fit individual sectors introduces a further constraint in our selection of systems due to the relatively short duration ($\sim 27$ days) of a sector. In order to ensure a good fitting of the transits, we decide that at least five transits should be present in the light curve captured in each sector. This effectively results in a restriction on planetary orbital periods, which are needed to be shorter than five days. With this final filtering of the list, we are left with a selection of 46 TESS confirmed planets in systems with a single star and a single planet, with more than one cadence available and with periods less than five days. These 46 systems translate to a total of 556 single-sector light curves to be fitted. A list of all the selected systems as well as the cadences available for each cadence is provided in the GitHub repository\footnote{\url{https://github.com/JulioHC00/PEPITA}}.

\subsection{Light curve processing and fitting}
\label{sub:light curve_pre_processing_and_fitting}

\begin{table}
\caption{Transit and GP variables fitted with the Bayesian model and their priors. $\mathcal{N}(\mu, \sigma^{2})$ represents a normal distribution with mean $\mu$ and standard deviation $\sigma$, $\lvert \mathcal{N} \rvert(\mu,\sigma^{2})$, represents a normal distribution of only positive values with mean $\mu$ and standard deviation $\sigma$ and $\mathcal{U}(a,b)$ represents a uniform distribution between $a$ and $b$. Values of the form $\textsc{nea} x$ indicate that the value of $x$ comes from the NEA table. Similarly, a standard deviation of the form $\textsc{nea} \sigma_{x}$ indicates that the standard deviation is taken as the error reported in the NEA table value. Fallback values for the period and transit time standard deviations in case no value is available in the NEA table are $10^{-3}$.}
\label{tab:transit_variables}
\centering
  \begin{tabular}{l r}
	  \hline
	  Variable & Prior \\
	  \hline
	  $P$ & $\mathcal{N}$(\textsc{nea} $P$, \textsc{nea} $\sigma_{P}^{2}$) \\
	  $t_0$ & $\mathcal{N}$(\textsc{nea} $t_0$, \textsc{nea} $\sigma_{t_0}^{2}$) \\
	  $\log{R_{P}/R_{*}}$ & $\mathcal{N}$($\sqrt{\text{bls\_depth}}$, $5^{2}$) \\
	  $b$ & $\mathcal{U}$(0,1) \\
	  $\{u_1, u_2\}$ & Uninformative, as described in \cite{exoplanet:kipping13} \\
	  $\hat{F}$ & $\mathcal{N}$(0, $1^{2}$) \\
	  $M_{*}$ & $\lvert{\mathcal{N}\rvert}$(\textsc{nea} $M_*$, $10^{2}$) \\
	  $R_{*}$ & $\lvert{\mathcal{N}\rvert}$(\textsc{nea} $M_*$, \textsc{nea} $\sigma_{R_*}^{2}$) \\
	  \hline
	  $\log{\sigma}$ & $\mathcal{N}$(0, $0.1^{2}$) \\
	  $\log{\rho}$ & $\mathcal{N}$(0, $0.1^{2}$) \\
	  \hline
  \end{tabular}
\end{table}

Pre-processing of the light curve starts by applying a Savitzky-Golay filter \citep{savitzkySmoothingDifferentiationData2002} to the light curve in order to remove any present trends in the data---vibration of the instrument or stellar variability, among others. In order to prevent any deformation in the transits by the application of the filter, two precautions are taken:

\begin{enumerate}
  \item A transit mask is created using the transit duration, period, and transit time values from the NEA table. This mask is used to prevent the filter from incorporating the transits into the calculation of trends and protects the transits from deformations.
  \item The window length of the filter is chosen to be five times the number of data points in a single transit to further discourage the filtering of trends associated with the transits.
\end{enumerate}

Additionally, periodograms are calculated for the filtered light curves using a box least squares procedure from which values for the period, transit time and depth of the transit (bls\_depth) are obtained. The square root of the value of the transit depth is used as a starting guess of the planet radius ratio, but the period and transit time obtained from the box least squares are only used as wherever no values for these quantities are available in the NEA table.

The final pre-processing step consists of modelling any remaining distortions in the data by using a Gaussian process, which finds functions that predict trends in the data \citep[for more details about Gaussian processes and their application to exoplanet transits see][]{barrosImprovingTransitCharacterisation2020} and outlier identification after the distortion has been removed. To do this, a Bayesian model is set up using the \texttt{exoplanet} package \citep{exoplanet:joss, exoplanet:zenodo} and \texttt{PyMC3} \citep{pymc3} while the Gaussian process (GP) is implemented using \texttt{celerite2} \citep{celerite1, celerite2}. The transit is modelled using the parameters $\{P, t_0, \log{R_{P}/R_{*}}, b, u_1, u_2, \hat{F}, M_{*}, R_{*}\}$ where $P$ is the orbital period of the planet, $t_0$ is a reference time for the transit (the mid-transit time of a reference transit), $R_p/R_*$ is the radius ratio between the planet and the star, $b$ is the impact parameter, $\{u_1, u_2\}$ are the parameters for a quadratic limb-darkening model, $\hat{F}$ is the mean value of the flux in the light curve, $M_*$ is the mass of the host star and $R_*$ is the radius of the host star. Meanwhile, for the GP a SHOTerm\footnote{\href{https://celerite2.readthedocs.io/en/latest/api/python/?highlight=SHOTerm\#celerite2.terms.SHOTerm}{celerite2 documentation}} is used as the kernel and its parameters $\left\{\log{\sigma}, \log{\rho}\right\}$ (see documentation for the meaning of these parameters) are fitted simultaneously with the transit to model any trends in the residuals. The procedure here follows the basic structure described in the TESS case study presented in the \texttt{exoplanet} package\footnote{\href{https://gallery.exoplanet.codes/tutorials/tess/\#the-transit-model-in-pymc3}{\texttt{exoplanet} documentation}}.

Tight priors are only placed on the period, transit time and radius of the star, with the rest of the variables set with either loose Gaussian priors or uniform priors. This is to allow the information contained within each light curve to determine the shape of the posteriors, rather than it be dominated by a tight prior, so that any difference between cadences becomes apparent. As a note of caution, we chose not to fit for eccentricity for computational efficiency given the large sample of light curves to be fitted. Any signatures left behind in the light curve should be absorbed into the posterior of the stellar density \citep[see e.g.][]{eylenOrbitalEccentricitySmall2019, dawsonPhotoeccentricEffectProtohot2012} so we expect any effect on the radius ratio determination to be negligible. Additionally, a uniform prior between 0 and 1 is placed on the impact parameter, meaning that grazing transits---those for which $b>1$---can not be modelled. This should not be an issue as there are no planets with grazing transits in the selected sample of systems according to $b$ values from the NEA table. For the GP, the parameters have tight priors to prevent overfitting. In Table \ref{tab:transit_variables} all the transit variables and their priors are summarized.

With the model setup, we obtain the first maximum likelihood or best fitting set of parameters. Using these parameters, we subtract the best fitting light curve and the GP to the data and then remove any point whose residual is larger than 5 times the root-mean-square value of all residuals. With the outliers removed, we obtain a new second set of best fitting parameters, which are then used as a starting point for the Markov Chain Monte Carlo sampling.

Finally, we sample two chains with 8000 tuning samples and 8000 normal samples, with the starting point given by the best fitting parameters of the light curve without outliers. The target accept parameter defines the target acceptance ratio of our MCMC. The acceptance ratio is defined as the ratio between the number of accepted samples and the total number of samples. We choose a value of 0.97 through experimentation as we found it produced good fits. Seeds for each chain are generated randomly. Convergence of the chains is checked through the ``rhat'' parameter, which must be close to 1 for convergence \citep{vehtariRanknormalizationFoldingLocalization2021}.

\subsection{Numerical Information Analysis}
\label{sub:numerical_analysis}

The Information matrix technique (Information Analysis henceforth) is a mathematical formalism that allows---under some conditions that the data must meet---predicting the best precision one can expect to obtain in your model parameters after conducting an experiment but without having to perform the experiment or having to simulate it in detail \citep{wittmanFisherMatrixBeginners2016}. It does this by ``measuring'' how much information is contained in the combination of the data and the model. Returning to Figure \ref{fig:two_cadences.pdf} we can understand that the 20s cadence model is more sensitive to changes in the input parameters while the 1800s model has lost some of that sensitivity. That is, small changes in the input parameters produce larger changes in the produced light curve with 20s cadence than with 1800s cadence. We could then expect that 20s data should produce more precise values since small deviations from the true parameters will quickly make the predicted light curve deviate from the data, while 1800s will produce worse precisions. The Information Analysis formalizes this intuitive idea by introducing the concept of information, plus it takes into account the errors in our measurements. More than that, the information analysis technique also allows you to predict covariances between parameters, potentially warning you of the need for reparametrization for a more efficient fitting of your parameters. For a formal description of the Information matrix formalism, see \cite{kaganRelationCovarianceFisher1999}.

The condition that our data must fulfil in order for this analysis to be valid is that errors in the data points must be Gaussian with a mean of 0 and \emph{uncorrelated}. Here, we will assume that this condition is satisfied by our data.

To perform the analysis itself, we define $F\{t_{k};\{p_{m}\}\}$ as the transit flux model, which is evaluated at points in time $t_{k}$ and which depends on a set of parameters $\{p_{m}\}$. The standard deviation in the point at time $k$ is taken to be $\sigma_{k}$ and then we can calculate the entries in the zero-mean Gaussian-noise Information matrix using the following expression

\begin{equation}
  B_{ij} = \sum_{k=1}^{N} \sum_{l=1}^{N}\left[ \pdv{}{p_{i}}F\left( t_{k};{p_{m}} \right) \right]\mathcal{B}_{kl}\left[ \pdv{}{p_{j}}F\left( t_{l};{p_{m}} \right) \right],
    \label{eq:fish}
\end{equation}
\noindent where the derivatives are with respect to model parameters, and we sum over all times. Here, $\mathcal{B}_{kl}$ is the inverse of the covariance matrix of the measurements, which in this case will be just a diagonal matrix with $\sigma^{-2}_{k}$ in the diagonal. That is, $\mathcal{B}_{kl} = \delta_{kl}\sigma_{k}^{-2}$. Then, with the conditions described above met, we can obtain the elements of the covariance matrix by simply inverting this matrix

\begin{equation}
 \text{Cov}(p_{i},p_{j}) = (B^{-1})_{ij}.
\end{equation}

The diagonal of the matrix will give us the smallest possible variance we can expect on each of the parameters, while off-diagonal elements measure the covariance between the parameters. Of course, this is a prediction of the best precision we can obtain in the parameters, and many factors can result in our experiment obtaining larger variances.

Thus, all the difficulty of the analysis lies in calculating the derivatives of our flux model, which are not guaranteed to be analytically derivable. Calculating analytical derivatives of limb-darkened flux models is not possible, as binned light curves have to be calculated numerically, and calculating numerical derivatives can be expensive computationally. Previous works which implemented this technique in the context of exoplanet transits \citep{carterAnalyticApproximationsTransit2008, priceTransitlightcurves2014} used simplified, linear trapezoidal transit models with no limb-darkening in order to be able to derive analytical expressions in the interest of computational efficiency. Their Information Analysis predictions of the variances agree well with MCMCs run on artificially-generated data based on the linear trapezoidal model, showing the validity of the analysis. However, the introduction of limb-darkening makes the predictions less accurate and the limitations go beyond the lack of limb-darkening effects in the model. The parameters for which they derive variances are related to the morphology of the light curve (e.g. duration of the ingress, depth of the transit or duration of the full transit among others) which are related in complex ways to the more physical parameters normally used to describe a transit (e.g. radius ratio, stellar density or impact parameter among others) and the data points---the times at which observations were made---have to be assumed to be perfectly uniformly sampled which is not necessarily the case as interruptions are frequent in observations. 

In order to address the limitations of an analytical approach to the Information Analysis technique, we propose a fast numerical implementation such that the \emph{exact model} that is going to be fitted is used directly in the calculation of the matrix---allowing the prediction of the precision of more physical variables such as the impact parameter and the density of the host star. Moreover, a numerical analysis allows for the prediction of the variances associated with the exact distribution in time of data points available, including interruptions or any other deviation from a uniform sampling of data points. We believe that our approach increases the appeal of the technique as it moves it closer to the hands-on work of observations and away from the theoretical realm, making it more suitable to be used for the efficient planning of observations of exoplanets whose parameters want to be refined.

\subsubsection{Implementation}
\label{ssub:implementation}

In order to efficiently  implement the numerical Information Analysis of transit light curves, we make use again of the \texttt{exoplanet} package. The implementation of the package with \texttt{theano}\citep{theanodevelopmentteamTheanoPythonFramework2016b}, allows performing the numerical derivatives required for the computation of the matrix and do so in a fast an efficient manner (Dan Foreman-Mackey, private communication).

Thus, Prediction of Exoplanet Precisions using Information in Transit Analysis (\texttt{PEPITA}) was developed by implementing the transit model described before and including methods were included for the calculation of the derivatives, the Information matrix and the covariance matrix. The implementation is designed in such a way that modifications of the particular transit model are easy to implement. All that is needed for the calculation of the covariance matrix is to provide an array of timestamps of when data points were collected along with the errors in the measurement of each of the data points (in our case taken directly as reported in the SPOC light curve) or a mean error that is assumed to be equal for all points, as well as a set of parameters defining the fiducial model---the model used to evaluate the derivatives and usually the best-fit set of parameters. In principle, the derived covariance matrix should  make a good prediction regardless of the chosen fiducial model, as long as the real parameters are not too far from them.

\begin{figure*}
  \centering
  \includegraphics[width=1.0\textwidth]{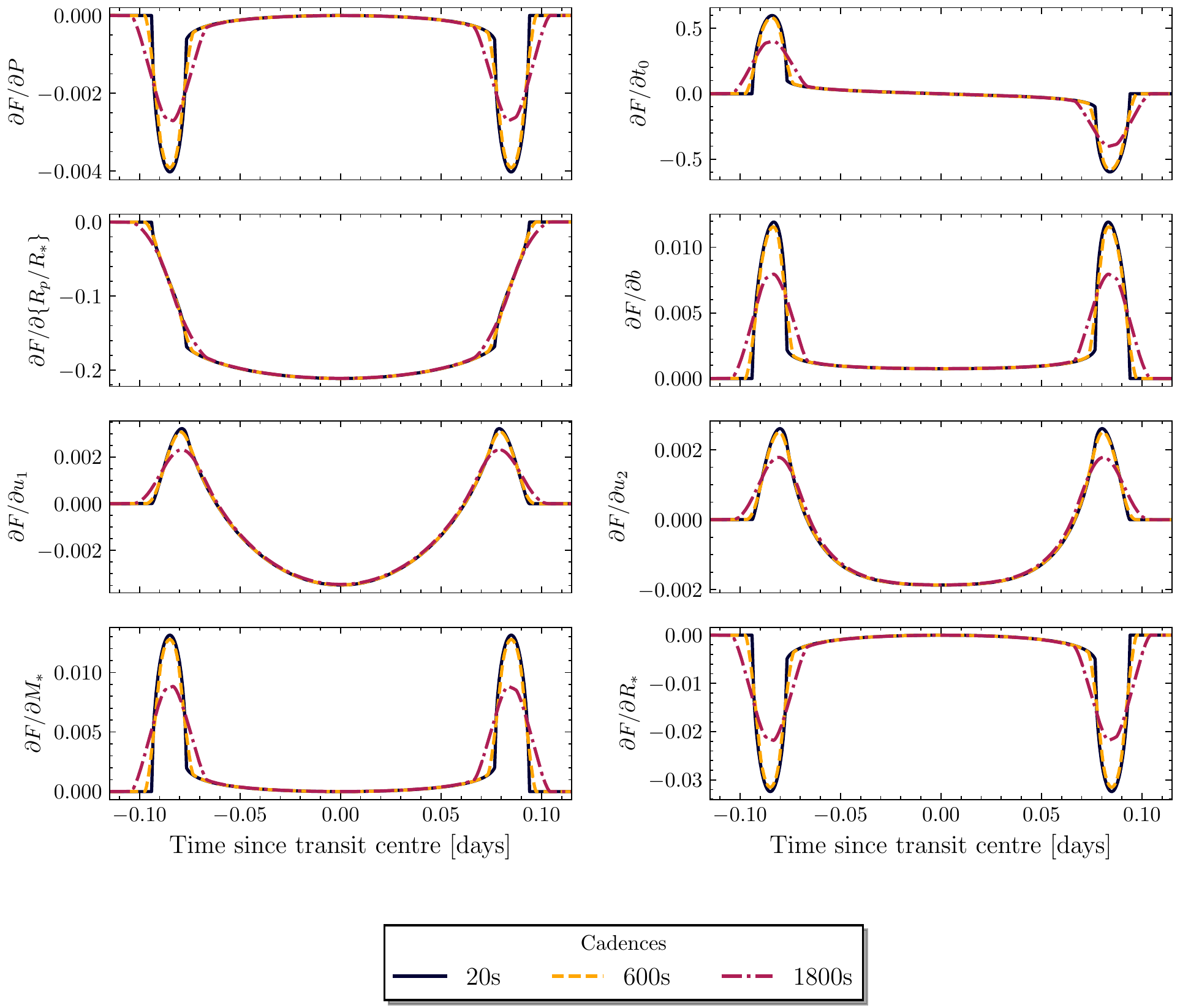}
  \caption{Derivatives of the transit light curve of HD~2685 with respect to model parameters. The derivatives are computed for cadences of 20s (black), 600s (orange) and 1800s (magenta) using the integration of the \texttt{exoplanet} package with \texttt{theano}. Just from the derivatives, it is intuitive to see that a 1800s light curve is less sensitive to changes in the parameters and will do worse in constraining them, while 600s and 20s almost as equally sensitive and should produce better precisions.}
  \label{fig:derivs.pdf} 
\end{figure*}

In Figure \ref{fig:derivs.pdf} we show the numerical derivatives calculated for HD~2685 with 20s (black), 600s (orange) and 1800s (magenta) cadences. For clarity, we did not include 120s derivatives, since they are very similar to 20s derivatives. Just by a visual inspection of these derivatives, one can gain an understanding of why longer cadences are, in principle, worse performing in the constraining of transit parameters. The 1800s derivatives are smaller than the 20s derivatives, which indicates that observations made in 1800s will be less sensitive to small changes in the model parameters and the precision derived from them will be worse than what can be obtained with shorter cadences. Of course, the exact difference between different cadences will depend on the set of parameters used to calculate the derivatives. In other words, 600s derivatives have almost the same shape as 20s derivatives for this particular set of parameters, but this will not necessarily happen for other planets. One should also remember that the Information matrix consists of not only the derivatives, but also the precision (standard deviation) of the data. The magnitude of the error bars scales as $\propto 1/\sqrt{\mathcal{I}}$ with $\mathcal{I}$, the cadence of the observations, and so while shorter cadence models will be more sensitive, they will also suffer from larger error bars in the data. Therefore, it is not possible to say that one cadence may be better than another without performing the analysis.

Another important takeaway from Figure \ref{fig:derivs.pdf} is noticing how some parameters produce derivatives that are an order of magnitude or larger than others or that are non-zero for wider ranges of time---compare, for example, the radius ratio and the impact parameter---which also explains why it is harder to constrain some parameters compared to others.

Priors are an important part of any Bayesian model that can be used to extract information from data, and these can too be incorporated into the Information analysis as another source of information that is independent on how much information is contained in the model. To do so, one must simply put these priors---the corresponding standard deviation of whatever prior distribution is being used---in their corresponding position of the diagonal of a new ``priors'' matrix which is then inverted so that the elements of the inverted matrix are given by

\begin{equation}
	(M^{-1})_{i, j} = \delta_{i, j}\sigma_{P_{i, j}}^{-2}
\end{equation}

with $\sigma_{P_{n,n}}$ the standard deviation of the prior distribution of parameter $n$. Note that having no prior would be equivalent to an infinitely large standard deviation, and so for that particular parameter the value of its entry in the inverted ``priors'' matrix will just be 0. That is, having no priors is the same as having no additional prior information.

All that is left to do is to add this matrix to the original Information matrix, and one has a matrix describing both the information contained in the data plus any information conveyed by our prior knowledge of the data. The inverse of this matrix provides us, just as before, with the covariance matrix whose diagonal elements will be the variances we expect to obtain by performing the MCMC fit.

\begin{equation}
    \text{Cov}(p_{i},p_{j}) = \left(\left[ B + M^{-1}\right] ^{-1}\right)_{ij}
\end{equation}

For the purpose of this work, we use the median values of the posteriors obtained from the MCMC fit for the fiducial model.

For the priors, we use the standard deviations described in Table \ref{tab:transit_variables} for Gaussian priors. For the impact parameter, we use $1/\sqrt{12}$, the standard deviation of a uniform distribution between 0 and 1. For the limb-darkening variables, although the posterior distributions that we obtain from the fitting are for $u_{1}$ and $u_{2}$, behind scenes the \texttt{exoplanet} package actually samples the reparametrizations described in \cite{exoplanet:kipping13}. Thus, even if the MCMC fit isn't able to constrain at all the values of these reparametrizations, it will still produce  non-uniform distributions for $u_{1}$ and $u_{2}$. These distributions correspond to a lack of extra knowledge derived from the data and so are just the prior on $u_{1}$ and $u_{2}$ that results from the reparametrization of these variables. Hence, we use the standard deviation of those distributions as the priors on $u_{1}$ and $u_{2}$.

\section{Results}
\label{sec:results}

\subsection{MCMC fit results}
\label{sub:mCMC_fit_resuts}

\begin{figure*}
   \centering
   \includegraphics[width=0.95\textwidth]{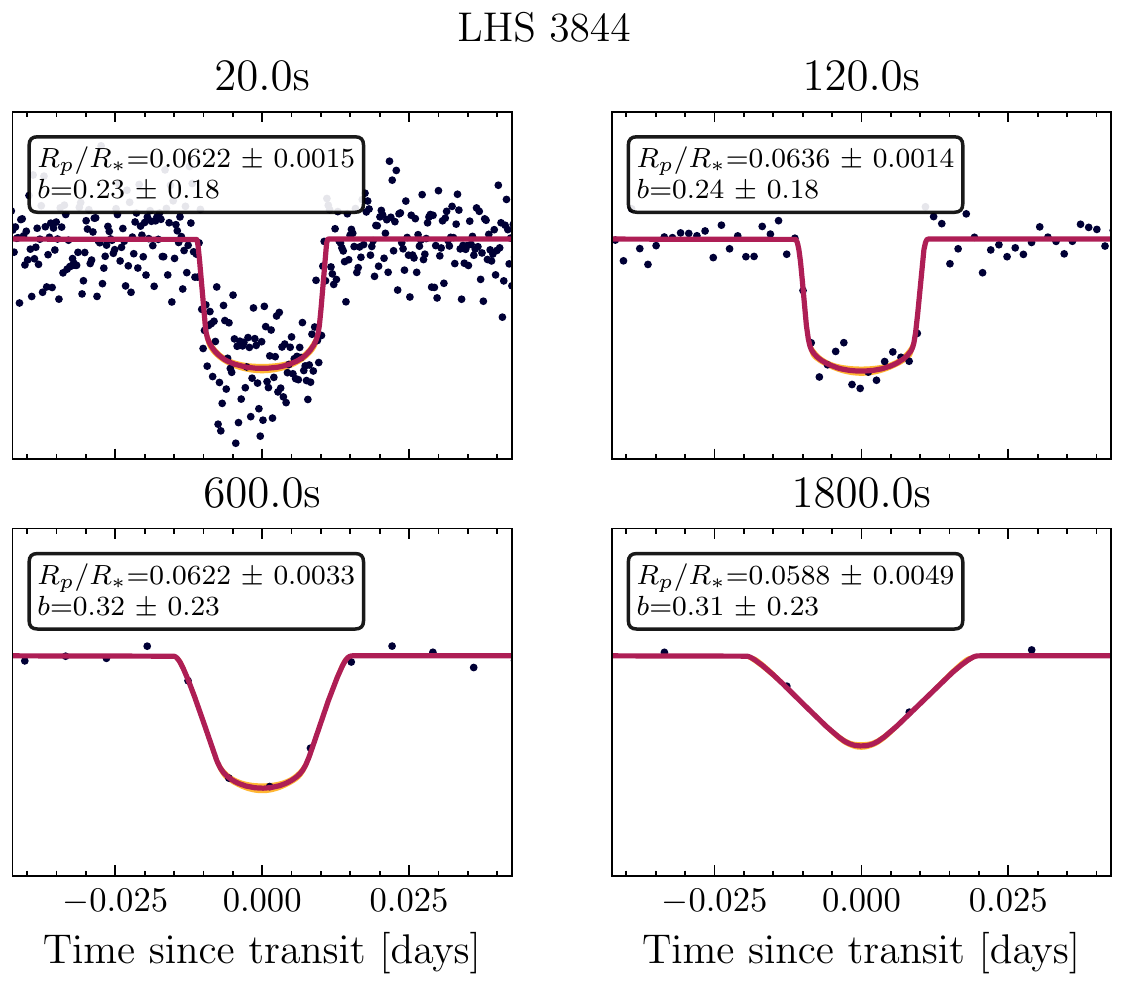}
   \caption{Final fits for LHS~3844 with median values and standard deviations of the posterior distributions of the radius ratio ($R_p/R_*$) and impact parameter ($b$) for each fit. To avoid cluttering by the data points, we fold light curves and then bin the data to the original cadence. That is, a 20s light curve will be folded, and then the folded data will be binned into 20s bins so that there is a single data point every 20s. From the posteriors, 10000 random samples are drawn for the transit parameters and light curves generated. The median of all samples is shown with a red line, while the range between the 16 and 84 percentiles is shown with a yellow coloured region. Individual plots like these were generated for each of the fitted light curves and manually inspected to identify any issues with the final fit before approving any results.}
   \label{fig:LHS_3844.pdf} 
 \end{figure*}

\begin{figure}
     \centering
     \includegraphics[width=1\columnwidth]{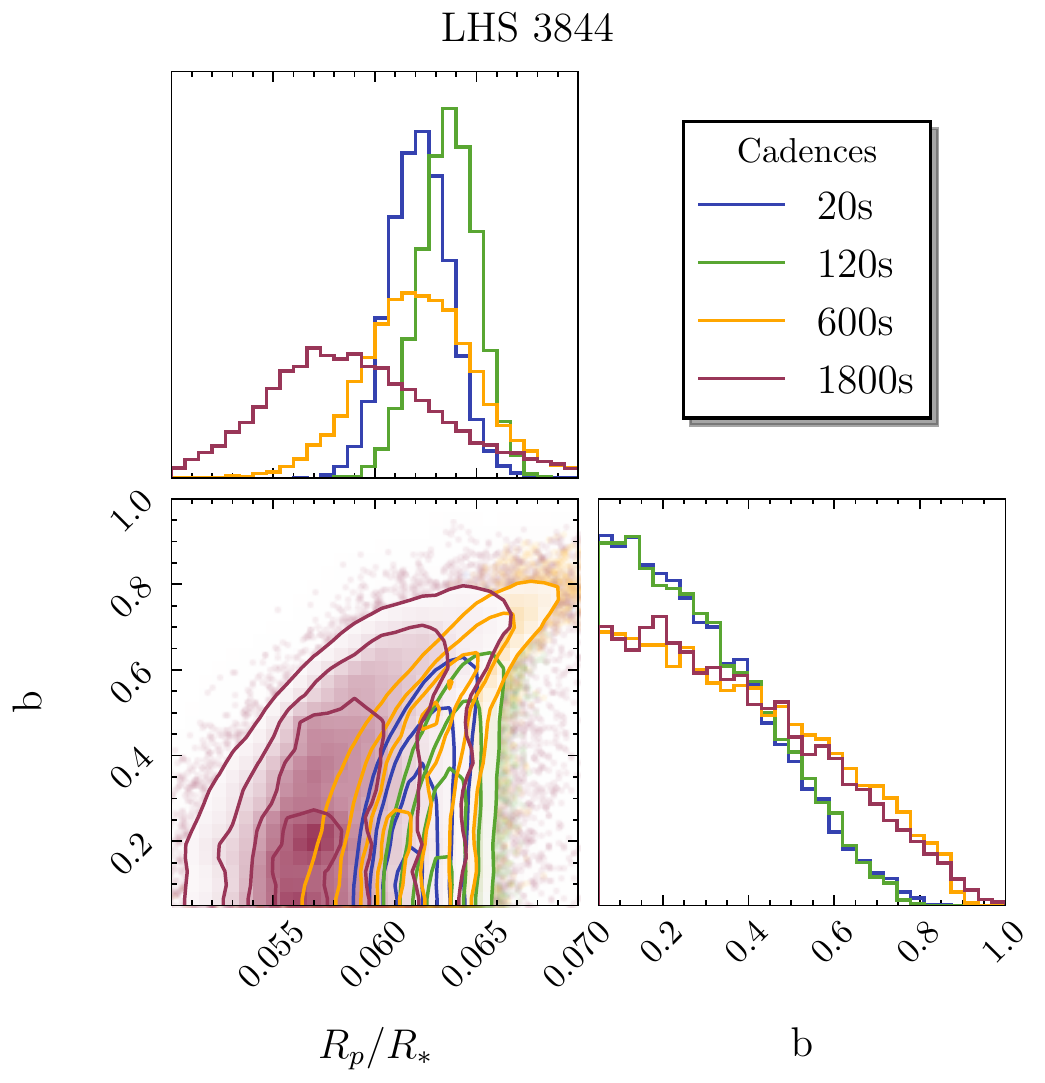}
     \caption{Cornerplot of $R_p/R_*$ and $b$ parameters for fits with cadences 20s, 120s, 600s and 1800s of LHS 3844. (c.f. Figure \ref{fig:LHS_3844.pdf}).}
     \label{fig:LHS_3844_corner}
 \end{figure}

Upon successful completion of each of the MCMC fits, a series of manual inspections were performed on each of the fitted light curves before the results were approved. Light curves were checked before and after the application of the Savitzky-Golay filter to ensure the transit was not eliminated or distorted by the application of the filter. Points marked as outliers and removed are also checked for each of the light curves to ensure that the pre-processing best-fit model and GP have not failed and caused valid points in the transit to be flagged as outliers. Additionally, a plot is made where the folded light curve is shown before and after the application of the median GP obtained after sampling is finished. This is done to ensure that there has not been an overfitting of the transit by the GP. For the last two checks, we generate a folded light curve, where the model light curves obtained by choosing 10000 random samples of the posteriors are plotted with the median shown as a line and the range between the 16 and 84 percentiles shown with a color-filled region (see Figure \ref{fig:LHS_3844.pdf}) as well as a cornerplot to visually inspect the posteriors (see Figure \ref{fig:LHS_3844_corner}, original cornerplots contain all variables while Figure \ref{fig:LHS_3844_corner} is a scaled-down version for clarity). In particular, Figures \ref{fig:LHS_3844.pdf} and \ref{fig:LHS_3844_corner} show how for this planet, good fits are obtained for all cadences and fits to the shorter cadence produce a higher precision in terms of their radius ratio and impact parameter.

Given that, as mentioned before, there may be several sectors for a particular system-cadence combination, there are several choices available when deciding how to compare the precision obtained by the different cadences of a same system. Ideally, we would compare the performance of the same sector fitted with different cadences. However, this choice would drastically reduce the number of comparisons possible, sometimes even making comparisons between different cadences impossible. As such, we choose to compare all possible combinations between a particular system-cadence-sector combination and all other system-cadence-sector combinations with a different cadence. That way, the number of comparison becomes large (more than 6000 comparisons in total), and although some sectors of a particular system-cadence combination may have performed slightly better or worse than others, the deviation should average out as we compare everything with everything.

\begin{figure*}
  \centering
  \includegraphics[width=0.95\textwidth]{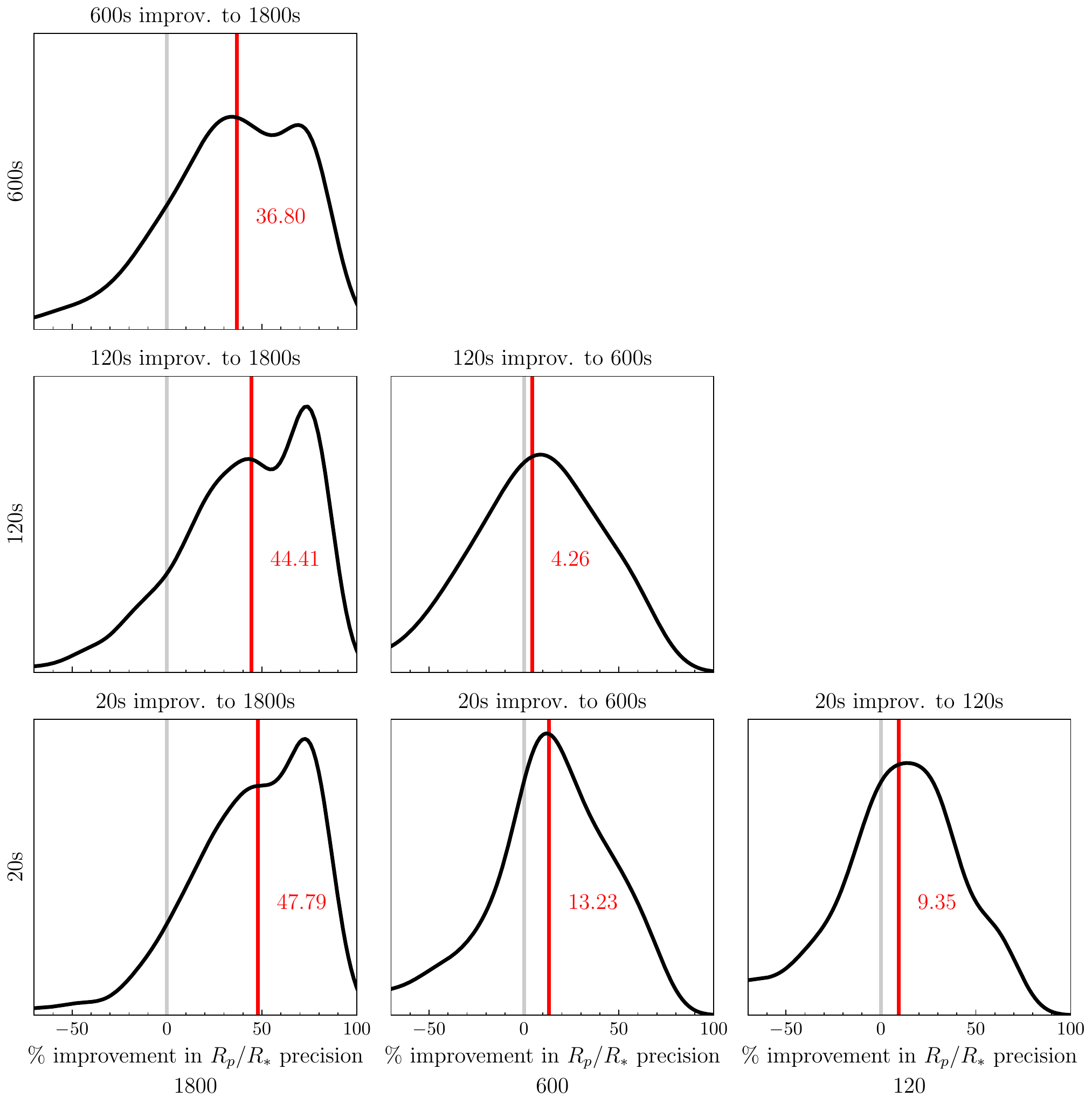}
  \caption{Density plot of the distribution of improvements (see Equation \ref{eq:improvement}) observed for every combination of cadences. Values above 0 indicate an improvement of precision with the use of a shorter cadence, while values below 0 indicate a worsening of the precision. Vertical grey lines indicate neither an improvement nor a worsening at a value of $0\%$ while red lines indicate the median improvement observed for a particular cadence comparison.}
  \label{fig:ror_improv.pdf} 
\end{figure*}

In Figure \ref{fig:ror_improv.pdf} we show the distribution of precision improvements observed for all possible cadence combinations out of the \{20s, 120s, 600s, 1800s\} cadences available for TESS observations. We calculate the precision improvements as follows. Given $\sigma_y$ the standard deviation of the posterior fit for the radius ratio of the shorter cadence (y-axis label in the plot) and $\sigma_x$ that of the longer cadence (x-axis label in the plot) the improvement is given by

\begin{equation}
    \text{Improv.} = \left(1 - \frac{\sigma_y}{\sigma_x}\right) \cdot 100
    \label{eq:improvement}
\end{equation}

so that a positive value of 50 means that the shorter cadence performed $50\%$ better than the longer cadence.

\subsection{Information analysis results}

\begin{figure*}
    \centering
    \includegraphics[width=0.95\textwidth]{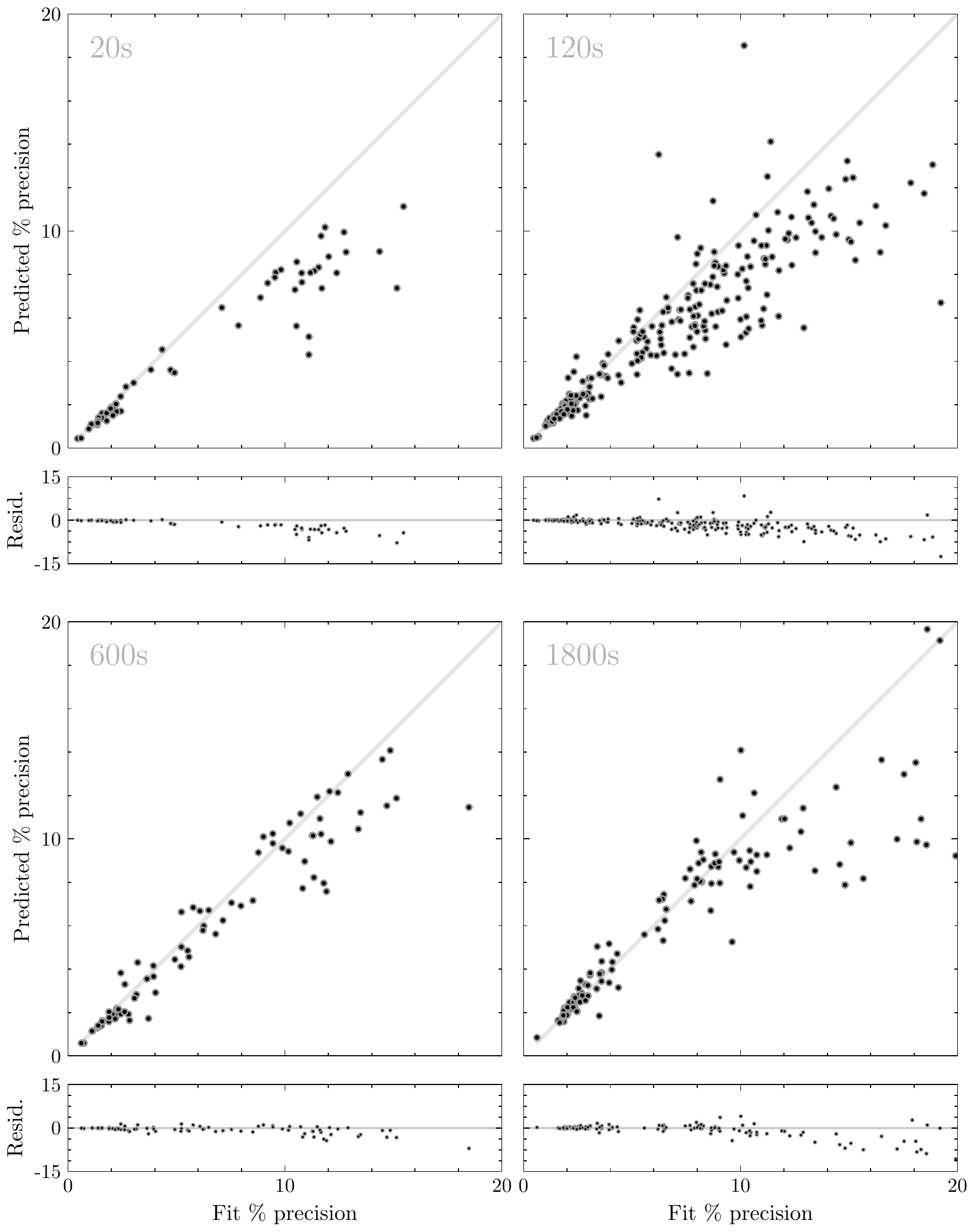}
    \caption{Predicted $\%$ precisions obtained using the Information analysis against precisions obtained from the MCMC fits. Points on the grey diagonal line indicate a perfect agreement between predicted and real value. Meanwhile, values above the line indicate an underprediction of the precision, while values below indicate an overprediction. Residual plots are included, with the difference between the predicted precision and the fit precisions plotted against fit precisions.}
    \label{fig:information_results.pdf}
\end{figure*}

In order to display the results of the Information analysis predictions, we compare the precision as obtained from the MCMC fit to that predicted with the Information analysis. We show in Figure \ref{fig:information_results.pdf} the predicted $\%$ precisions against the MCMC fit precisions, as well as the residual obtained by subtracting the fit values to the predicted values.

\subsection{Multisector results}

\begin{figure}
    \centering
    \includegraphics[width=1\columnwidth]{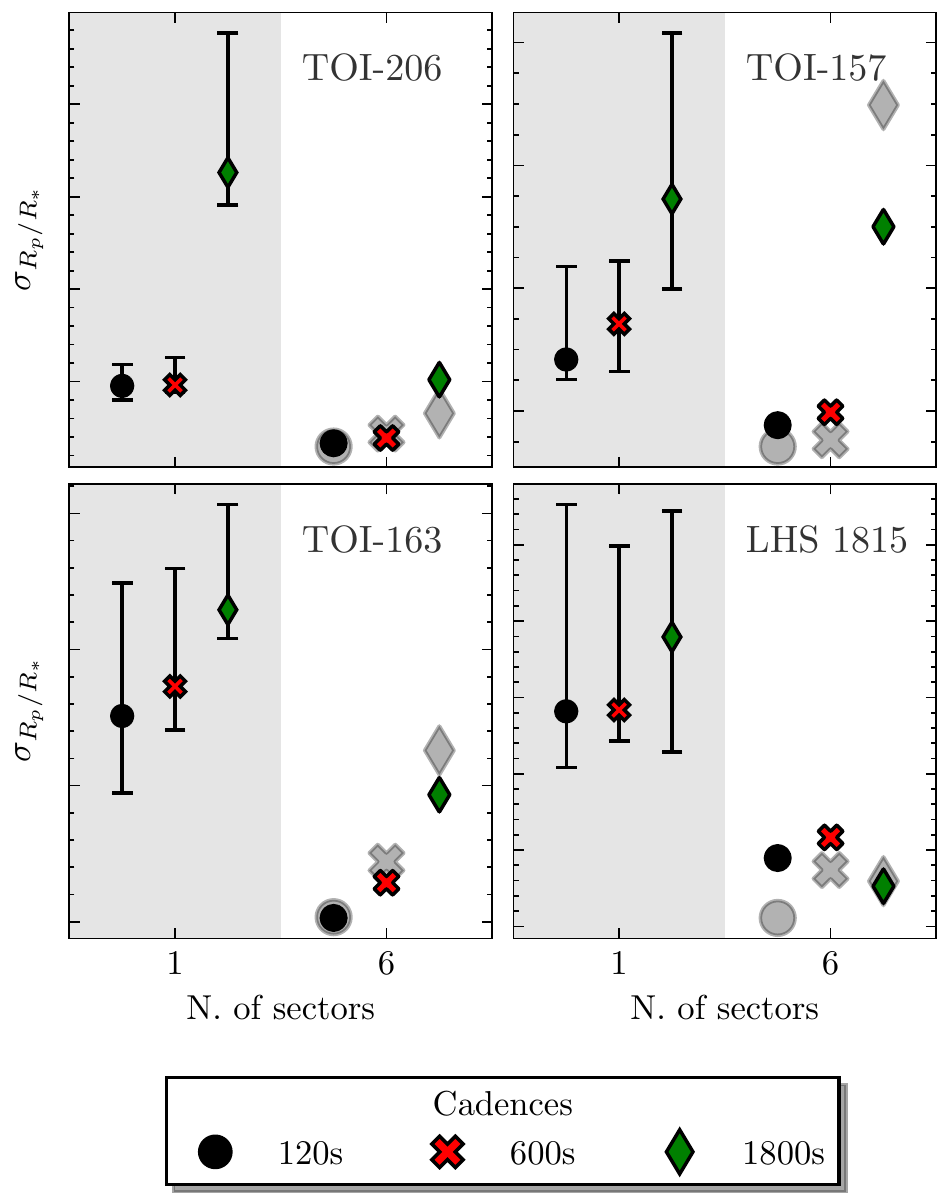}
    \caption{Precisions obtained from fit to a single sector and to 6 sectors simultaneously for 4 planets from our selection. Single sector values correspond to the mean precision of the individual fits to each of the 6 sectors used in the multisector fit, and error bars indicate the 16 and 84 percentiles. The grey, larger markers on the 6 sector fits show the precision predicted by the Information analysis applied to the 6 sectors simultaneously.}
    \label{fig:multisector.pdf}
\end{figure}

Since fits for published parameter values are usually performed to more than one sector at a time, we perform multisector fits for a small number of systems from our selection. In Figure \ref{fig:multisector.pdf}, we show the MCMC fit precisions obtained for a fit to a single sector as well as for a fit to 6 sectors simultaneously for 4 systems of our selection. We also plot the Information analysis predicted precisions for the 6 sector fits. 

We find that it is hard to tell how a multisector fit will perform exactly based on the single sector precisions, besides the expectation that precisions should increase with the use of more sectors. Nevertheless, we found no hints of the inclusion of several sectors at the same time in the Information analysis affecting its performance.

\subsection{Predictions for TESS Objects of Interest (TOIs)} 
\label{sub:predictions_for_tESS_objects_of_interest_tOI_}

Given the encouraging results of the Information analysis, we provide as a proof of concept, the predicted radius ratio improvements to be obtained by the reobservation of TOIs with different cadences.

To perform this analysis, we start with the TOIs table from the NASA Exoplanet Archive and select only objects of interest which are identified as planet candidates. Then, only objects with values for $\log{g}$, $R_*$ and $T_{\text{eff}}$ (stellar effective temperature) along with errors available in the table are chosen. From these, we select only those planet candidates \emph{for which only 1800s observations are available}. The values we extract from the table for each of the TOIs are the transit duration $T$, the transit depth $\delta$, period, transit time and the stellar radius, surface gravity and effective temperature.

The analysis is performed in much a similar way to what was described above. However, in this case, we choose to set up the Information Analysis by modelling the transit using the log of stellar density, $\log{\rho_{*}}$, instead of the stellar mass (with no prior on the stellar density). Because values for the stellar density, the impact parameter and the stellar limb-darkening parameter are needed in order to construct the fiducial model, we obtain these values by combining the available parameters as follows.

The limb-darkening parameters are obtained using the \texttt{Python Limb Darkening Toolkit} or \texttt{PyLDTk} \citep{Parviainen2015, Husser2013}. Since no value for stellar metallicity is available, we use $z=0.25$ with error 0.125 for all the TOIs. Although the package can also provide errors in the calculated limb-darkening values, we still choose to use the same priors in the limb-darkening parameters as we used in the Information Analysis of the MCMC fits. We choose to do this, because the calculations of these limb-darkening parameters are very crude given the available stellar parameters.

Meanwhile, the impact parameter of the fiducial model is calculated by assuming a circular orbit. Under this assumption, the transit duration ($T$) can be combined with the period ($P$), stellar gravity ($g_*$) and stellar radii ($R_*$) to obtain the impact parameter as

\begin{equation}
b = \sqrt{1 - \left(\frac{g_*\pi}{4 P R_*}\right)^{2/3} T^{2}}.
\end{equation}

Meanwhile, for the fiducial model value of $\rho_*$ we obtain an expression combining $g$ and $R_{*}$ by using Kepler's laws. With this, $\rho_*$ is given by

\begin{equation}
	\rho_* = \frac{3}{4 \pi G} \frac{g_*}{R_*}.
\end{equation}

Finally, for the radius ratio, once the limb-darkening parameters and the impact parameters are obtained we can get an approximate value for the ratio by using the transit depth in combination with the limb-darkening parameters and the impact parameter. This functionality is already implemented in the \texttt{exoplanet} package, and we make use of it.

With the fiducial model parameters all determined, we download the available 1800s cadence light curves for the TOIs. The array of timestamps is taken to be an array of times evenly spaced by each of the cadences between the minimum time from the downloaded 1800s light curve and the maximum time of the downloaded 1800s light curve. That is, if the first point in the 1800s cadence is at time $t$, we create a uniform array of points separated by, for example, 20s for the analysis of 20s cadence starting at $t$ and extending up to the last point of the 1800s cadence. This is to ensure homogeneity between cadences. For the errors in the observations, we take the mean error of the 1800s data points of each planet candidate $\bar{\sigma}_{1800\text{s}}$ and then for each of the other cadences of $x$ seconds the error ($\sigma_{x\text{s}}$) is assumed to be the same for all points and equal to
\begin{equation}
	\sigma_{x\text{s}} = \bar{\sigma}_{1800\text{s}} \sqrt{\frac{1800}{x}},
\end{equation}
\noindent this need for approximating the errors should be more carefully examinated if this analysis is repeated in a more rigorous manner as it can directly affect the predictions of the analysis.

Finally, we perform the Information Analysis for cadences 20s, 120s and 1800s, and we compare precisions in the radius ratio for each of the cadences.

Table \ref{tab:best10_toi} shows the expected improvements by reobservations with either 20s or 120s cadences for the 10 TOIs with the largest improvements. A full list is available in the GitHub repository\footnote{\url{https://github.com/JulioHC00/PEPITA}} along with a jupyter notebook showing the code used to make these predictions. The use of our Information Analysis implementation shown there can be extended to the analysis of other planet candidates/cadences.

\begin{table}
    \centering
\begin{tabular}{rrrr}
\hline
    TOI &  20s Improv. [\%] &  120s Improv [\%] &  600s Improv [\%] \\
\hline
1677.01 &            77.93 &            77.71 &            72.96 \\
2784.01 &            70.94 &            70.35 &            54.13 \\
3786.01 &            66.84 &            66.79 &            65.89 \\
1701.01 &            66.33 &            65.00 &            40.53 \\
2578.01 &            65.54 &            65.11 &            57.89 \\
4197.01 &            52.43 &            51.77 &            42.86 \\
5577.01 &            51.98 &            51.19 &            36.91 \\
5654.01 &            51.12 &            50.77 &            43.22 \\
3719.01 &            50.81 &            50.40 &            42.10 \\
2341.01 &            50.31 &            49.65 &            35.63 \\
\hline
\end{tabular}
    \caption{Expected improvements by reobservation with either 20s or 120s cadence of TOIs with only 1800s cadence observations. The 10 TOIs with the highest improvements are shown.}
    \label{tab:best10_toi}
\end{table}

\section{Discussion}
\label{sec:discussion}

The results presented in the previous section confirm our expectation that the precision of the radius ratio can be improved with the use of shorter cadences. We have observed an almost doubling (Improv.~$>~50\%$) of the precision in half of the comparisons between 20s and 120s cadences to 1800s cadences.

As expected, the median improvement observed is largest when comparing the shortest cadences to the longest cadences (i.e. 20s to 1800s).  The trends seen in the rows and columns of Figure \ref{fig:ror_improv.pdf} are consistent with the previous discussion about information being ``lost'' when a longer cadence is compared to a shorter one, and should encourage the use of shorter cadences wherever possible.

It is not as clear from Figure \ref{fig:ror_improv.pdf}, for example, whether 120s cadence observations are preferable to 600s with a median improvement of 4\%. However, it is worth emphasizing that the plots show the distribution of all comparisons between 120s and 600s cadence observations. The actual improvement will depend on the particular system being considered, as can be seen in Table \ref{tab:best10_toi}, where we present predicted improvements for 10 different TOIs and show that these improvements depend on the particular parameters of the system. For example, while TOI 3786.01 shows a very similar improvement in the radius ratio precision when reobserved with either 20s, 120s or 600s cadences (66.84\%, 66.79\% and 65.89\% respectively), TOI 1701.01 shows a clear difference between the improvement expected from reobservations with either 20s or 120s cadence of around 65\% and that expected from reobservations with 600 cadence of around 40\%.

This dependence of cadence effects in the obtained radius ratio precision on the particular system, highlights the need for a fast and easily adaptable prediction method, such as the one we present here and made publicly available. Figure \ref{fig:information_results.pdf} shows overall median errors in our predictions of about 1\%. We observe that, as the fit precision gets worse, the predictions start to deviate more and predict better precisions than the ones actually obtained in the fit. This can be understood by considering that, as the fit precisions get worse, it is possible that the fit is doing poorly for some reason our method did not account for. In those cases, since the Information Analysis predicts the best possible precision one should be able to obtain, our predictions are expected to be of better precisions than obtained with the fits. When fit precisions get worse than 10\% we find that median errors for 20s cadence predictions can reach values of around 10\% error. Therefore, predictions should be treated as an approximate lower boundary for the precision to be obtained with a fit. Whether this precision is obtained or not will depend on the fitting methodology. Where good fits are obtained, with precisions of a few \%, we expect predictions to be accurate up to $\lesssim 0.5\%$.

\begin{figure}
    \centering
    \includegraphics[width=0.99\columnwidth]{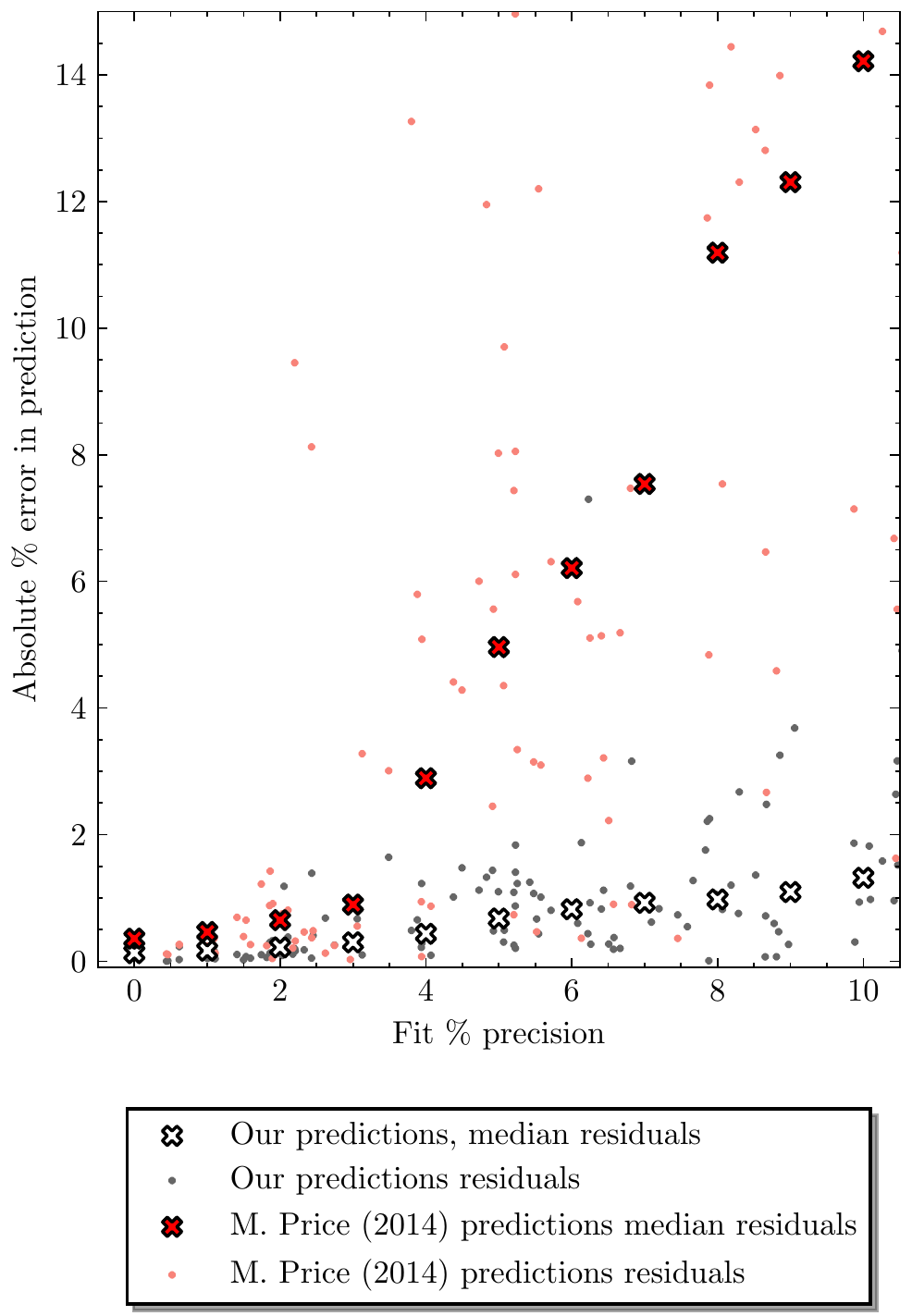}
    \caption{Residuals from our predictions (grey) and those obtained using the analytical methodology presented in \citet{priceTransitlightcurves2014} (red). Coloured crosses represent a moving median of the residuals, with a bin size of 6\% around the central value.}
    \label{fig:compare_price}
\end{figure}

When comparing our predictions (numerical method henceforth) to those obtained using the analytical methodology described in \citet{priceTransitlightcurves2014}  (analytical method henceforth), by obtaining predictions for the radius ratio standard deviation using their analysis and dividing it by the median value of the radius ratio, we find that our numerical method produces better predictions of the radius ratio precision, specially for larger values of \% precision. Moreover, while our methodology rarely predicts a precision worse than the precision obtained in the fit, analytical results suffer from this kind of error in most cases. This is relevant since, as discussed above, Information Analysis should predict the best attainable precision and thus be prone to predicting a precision better than the fit precision. That the analytical results suffer from underprediction errors is an indication that some of the information contained in the light curve failed to be captured by the analysis. In Figure \ref{fig:compare_price}, we show the absolute residuals from our predictions and the predictions obtained using the analytical approach when compared to fit precisions of the radius ratio. Coloured crosses represent the moving median of the residuals for a bin of size 6\% around the central value for our predictions (white) and analytical predictions (red). While the residuals from both prediction methods increase as the fit precision increases, those from an analytical prediction do so in a more pronounced manner. The code used to generate the predictions using the methodology of \mbox{\citet{priceTransitlightcurves2014}} (see equations A15 and A16 of their work) is available in the GitHub repository.

\section{Conclusion}
\label{sec:conclusion}

We have performed the uniform processing and fitting of 46 TESS confirmed planets transit light curves amounting to a total of 556 single-sector light curves to investigate the effects of cadence in the retrieved parameters. We model transits with the set of parameters $\{P, t_0, \log{R_{P}/R_{*}}, b, u_1, u_2, \hat{F}, M_{*}, R_{*}\}$ whose posterior distributions are obtained using a Markov Chain Monte Carlo (MCMC) procedure.

We also developed an implementation of a numerical Information Analysis technique to exoplanet transits by using the \texttt{exoplanet} package. \texttt{PEPITA} is highly adaptable to the exact set of parameters that are to be fitted and to the exact data points and precisions of the data points available. Thanks to the integration of \texttt{exoplanet} with \texttt{theano}, the required numerical derivatives can be obtained, making the numerical analysis possible and fast. Our numerical implementation differs from past analytical implementations of this technique in that it does not require approximations of the light curve shape and can be adapted to the exact model that is to be fitted while still producing fast results. This technique is applied to the light curves that were fitted with the MCMC.

Median improvements in the radius ratio precision of almost 50\% are observed when comparing fits to 20s or 120s cadence light curves to 1800s cadence light curves. Smaller, but still relevant, median improvements of around 35\% are also observed when comparing fits to 600s light curves to fits of 1800s light curves. However, it is important to highlight that these are median improvements only and that the actual improvement should be considered on a case-by-case basis. When we consider fits to multiple sectors simultaneously, we find no significant changes in the very limited sample considered.

With this in mind, we check the performance of our numerical information analysis by producing predictions of the radius ratio precision obtained by the fits for each of the light curves. While we find that for fits where the precision of the radius ratio is worse---that is, for large values of \% precision---our predictions perform worse, we speculate that this is likely caused by the poor performance of those fits due to our fitting model not accounting for certain factors which can affect the light curve. For example, removing stellar variability is done here with a general GP, but a more individualized analysis of each light curve could result in better removal of any variability. Another possible factor that is unaccounted in our model are stellar spots, which can also affect the light curve. Nevertheless, where the fit radius ratio precision is of a few \%, our predictions have errors $\lesssim 0.5 \%$. Even when larger errors of a few \% are observed, these correspond to predicting precisions better than observed, in line with how the information analysis should perform. When we compare our method to previous analytical methods, we find that not only are our errors significantly smaller but that the analytical predictions tend to be of worse precisions than observed, indicating that the analysis has missed some of the information contained in the light curve.

Given the satisfactory performance of our prediction method, we apply it to a number of TESS objects of interest with observations available only with 1800s cadence to demonstrate how our implementation of the information analysis can aid in deciding which targets should be prioritized for short cadence observations. We present the top 10 TOIs which would benefit most from reobservations with short cadences and highlight how for some of them there's an added benefit by using 20s or 120s cadences instead of 600s cadence while for others the difference is considerably smaller. Additionally, in Appendix \ref{appendix_full_list} we include the full list of all planet candidate predictions and make available the script used to make these predictions on the GitHub repository.

Our study has shown that shorter cadences offer better precisions in the radius ratio obtained from transits. Thus, whenever high precisions in this parameter are needed---constraining of the radius valley or the characterization of exoplanet atmospheres among others---special attention should be given to the choice of cadence as the choice of a shorter cadence is more probable to provide the required precision. We have not focused here in any of the particular areas that could benefit from the increased precisions, but we expect our results would be important in the context of investigations of the radius valley, transmission spectroscopy and the determination of exoplanet densities among others. Instead, our aim has been to highlight that there are indeed benefits to be gained by considering the use of shorter cadences. Moreover, we have demonstrated that \texttt{PEPITA} is able to accurately predict the precision in the radius ratio that can be obtained by fits to any particular light curve. Our implementation has the power of serving as a tool in the planning of future mission by providing information about which targets should be prioritized for observations with short cadences. In order to allow for such use \texttt{of PEPITA}, we have made it publicly available in GitHub\footnote{\url{https://github.com/JulioHC00/PEPITA}}. We encourage studies focused on one of those areas that require high precisions, to consider the issue of cadence in depth and perhaps to recommend reobservations of exoplanets of interest whenever the information analysis predicts that the required precision will be obtained.

\section*{Acknowledgements}

This research made use of 

\begin{itemize}
	\item \textbf{PyTransit} \citep{parviainenPyTransitFastEasy2015}
	\item \textbf{Lightkurve}, a Python package for Kepler and TESS data analysis \citep{2018ascl.soft12013L}.
	\item \textbf{Astropy},\footnote{http://www.astropy.org} a community-developed core Python package for Astronomy \citep{astropycollaborationAstropyCommunityPython2013b, astropycollaborationAstropyProjectBuilding2018b}.
	\item \textbf{astroquery} \citep{ginsburgAstroqueryAstronomicalWebquerying2019}
	\item \textbf{exoplanet} \citep{exoplanet:joss, exoplanet:zenodo} and its dependencies and other related tools: pymc3 \citep{pymc3}, theano \citep{theanodevelopmentteamTheanoPythonFramework2016b}, arviz \citep{exoplanet:joss}, starry \citep{exoplanet:luger18}, 
	\item \textbf{pandas} \citep{reback2020pandas, mckinney-proc-scipy-2010}
	\item \textbf{numpy} \citep{harris2020array}
	\item \textbf{matplotlib} \citep{Hunter:2007}
\end{itemize}

Additional thanks to Dr. Foreman-Mackey for providing an example jupyter-notebook with the implementation of numerical derivatives of transit light curves using the \texttt{exoplanet} package. Thanks also to the paper's reviewer for their helpful comments. JHC and CSKH would like to thank the Science and Technology Facilities Council (STFC) for funding support through PhD studentships.

This paper includes data collected by the TESS mission, which are publicly available from the Mikulski Archive for Space Telescopes (MAST). Funding for the TESS mission is provided by NASA’s Science Mission directorate.

\section*{Data Availability}

Results of the fits and the predictions made using the information analysis both for the fits and the TOIs are available in the GitHub repository at \url{https://github.com/JulioHC00/PEPITA}. The sourcecode of \texttt{PEPITA}, which allows making predictions using the information analysis, is also available in the repository along with example Jupyter Notebook.


\bibliographystyle{mnras}
\bibliography{citations}



\appendix

\newpage
\section{Full list of planet candidate predictions}
\label{appendix_full_list}

\begin{table}
\centering
\caption{Predictions for all the planet candidates considered in order of decreasing improvements to the radius ratio precision.}
\begin{tabular}{rrrr}
\hline
    TOI &  20s Improv. [\%] &  120s Improv [\%] &  600s Improv [\%] \\
\hline
1677.01 &            77.93 &            77.71 &            72.96 \\
2784.01 &            70.94 &            70.35 &            54.13 \\
3786.01 &            66.84 &            66.79 &            65.89 \\
1701.01 &            66.33 &            65.00 &            40.53 \\
2578.01 &            65.54 &            65.11 &            57.89 \\
4197.01 &            52.43 &            51.77 &            42.86 \\
5577.01 &            51.98 &            51.19 &            36.91 \\
5654.01 &            51.12 &            50.77 &            43.22 \\
3719.01 &            50.81 &            50.40 &            42.10 \\
2341.01 &            50.31 &            49.65 &            35.63 \\
3788.01 &            48.18 &            47.77 &            41.61 \\
3859.01 &            45.68 &            45.36 &            39.05 \\
1658.01 &            45.34 &            45.12 &            40.67 \\
3765.01 &            44.92 &            44.41 &            37.35 \\
3972.01 &            44.66 &            43.91 &            35.28 \\
3762.01 &            43.91 &            43.82 &            42.12 \\
3805.01 &            43.89 &            43.45 &            34.29 \\
3571.01 &            42.28 &            41.77 &            36.00 \\
5122.01 &            41.35 &            41.24 &            37.08 \\
3651.01 &            41.06 &            40.87 &            36.97 \\
3670.01 &            40.86 &            40.37 &            27.25 \\
5702.01 &            40.63 &            39.43 &            21.96 \\
5579.01 &            40.40 &            40.01 &            35.41 \\
3792.01 &            39.86 &            38.43 &            18.70 \\
3642.01 &            38.47 &            38.07 &            32.29 \\
3747.01 &            38.12 &            37.80 &            32.22 \\
2533.01 &            38.05 &            37.37 &            27.98 \\
3686.01 &            37.85 &            37.23 &            30.25 \\
3118.01 &            36.73 &            36.21 &            27.59 \\
3654.01 &            36.64 &            35.99 &            25.64 \\
1521.01 &            35.61 &            35.31 &            31.37 \\
3737.01 &            35.01 &            34.56 &            28.25 \\
5644.01 &            34.99 &            34.85 &            32.27 \\
5762.01 &            34.55 &            34.36 &            30.76 \\
1559.01 &            33.78 &            33.52 &            29.56 \\
1655.01 &            33.11 &            32.64 &            19.57 \\
3946.01 &            32.68 &            32.31 &            26.80 \\
3699.01 &            32.56 &            32.12 &            26.06 \\
2395.01 &            32.51 &            32.16 &            26.66 \\
3976.01 &            32.27 &            31.87 &            25.36 \\
 302.01 &            31.99 &            31.82 &            28.62 \\
5686.01 &            30.82 &            30.28 &            23.28 \\
3999.01 &            30.51 &            30.21 &            25.92 \\
3842.01 &            30.37 &            29.77 &            22.39 \\
2790.01 &            30.10 &            29.60 &            23.17 \\
3727.01 &            29.55 &            29.28 &            25.29 \\
3755.01 &            28.99 &            28.42 &            20.36 \\
4199.01 &            28.83 &            28.35 &            21.44 \\
3795.01 &            28.36 &            27.20 &            13.41 \\
5754.01 &            28.33 &            28.21 &            25.84 \\
3920.01 &            28.22 &            28.08 &            25.49 \\
3829.01 &            27.98 &            27.62 &            23.62 \\
2033.01 &            27.68 &            27.48 &            23.89 \\
4231.01 &            27.33 &            27.01 &            21.76 \\
\hline
\end{tabular}
\label{tab:long_toi}
\end{table}

\begin{table}
\centering
\contcaption{}
\begin{tabular}{rrrr}
\hline
    TOI &  20s Improv. [\%] &  120s Improv [\%] &  600s Improv [\%] \\
\hline
3763.01 &            26.47 &            25.90 &            18.67 \\
5656.01 &            26.26 &            25.21 &            12.57 \\
3573.01 &            26.16 &            26.02 &            23.13 \\
5669.01 &            26.03 &            25.60 &            19.59 \\
3703.01 &            25.85 &            25.62 &            21.88 \\
3660.01 &            24.22 &            24.06 &            21.01 \\
5458.01 &            23.95 &            23.76 &            21.17 \\
3773.01 &            23.89 &            23.73 &            20.69 \\
3671.01 &            23.69 &            23.53 &            20.19 \\
5482.01 &            23.44 &            23.28 &            20.05 \\
5479.01 &            23.18 &            23.03 &            21.51 \\
3335.01 &            22.66 &            22.15 &            14.94 \\
3800.01 &            22.55 &            22.26 &            18.25 \\
3664.01 &            22.35 &            22.06 &            17.47 \\
1536.01 &            22.21 &            22.14 &            20.91 \\
4204.01 &            22.18 &            21.72 &            17.34 \\
3980.01 &            22.17 &            22.07 &            20.09 \\
5773.01 &            21.68 &            21.25 &            15.42 \\
3768.01 &            21.61 &            21.27 &            16.62 \\
3640.01 &            21.34 &            21.20 &            18.72 \\
3645.01 &            21.16 &            20.68 &            14.60 \\
3769.01 &            21.09 &            20.88 &            17.27 \\
2350.01 &            21.04 &            20.24 &             6.64 \\
5749.01 &            20.72 &            20.45 &            16.05 \\
5459.01 &            20.65 &            20.52 &            17.96 \\
2036.01 &            20.57 &            20.39 &            17.81 \\
3744.01 &            20.13 &            19.99 &            17.60 \\
5681.01 &            19.99 &            19.64 &            11.13 \\
1605.01 &            19.92 &            19.78 &            17.28 \\
3733.01 &            19.72 &            19.60 &            17.39 \\
2060.01 &            19.65 &            19.57 &            18.05 \\
3721.01 &            19.29 &            19.04 &            16.00 \\
5618.01 &            18.86 &            18.67 &            15.94 \\
5647.01 &            18.79 &            18.67 &            16.39 \\
5486.01 &            18.47 &            18.09 &            13.29 \\
3690.01 &            18.46 &            18.31 &            15.87 \\
1551.01 &            18.16 &            17.91 &            15.09 \\
3754.01 &            18.03 &            17.72 &            13.39 \\
3772.01 &            17.93 &            17.56 &            14.26 \\
3802.01 &            17.74 &            17.55 &            15.15 \\
3679.01 &            17.53 &            17.26 &            13.72 \\
5345.01 &            17.49 &            15.56 &             3.36 \\
5427.01 &            17.11 &            16.83 &            12.44 \\
1596.01 &            16.96 &            16.89 &            15.48 \\
3663.01 &            16.81 &            16.78 &            16.05 \\
5426.01 &            16.78 &            15.83 &             5.68 \\
3899.01 &            16.63 &            16.46 &            14.52 \\
 981.01 &            16.49 &            14.26 &             5.87 \\
3718.01 &            16.40 &            16.05 &            11.30 \\
5646.01 &            15.85 &            14.94 &             5.61 \\
3296.01 &            15.79 &            15.61 &            12.88 \\
3652.01 &            15.50 &            15.43 &            14.22 \\
4074.01 &            15.48 &            15.25 &            12.27 \\
3698.01 &            15.26 &            15.05 &            12.60 \\
5764.01 &            15.05 &            14.93 &            12.74 \\
5468.01 &            14.39 &            14.33 &            12.98 \\
3732.01 &            14.19 &            13.89 &             9.65 \\
5592.01 &            13.72 &            13.52 &            11.03 \\
3688.01 &            13.57 &            13.41 &            11.12 \\
4646.01 &            13.47 &            13.25 &             9.29 \\
5355.01 &            13.17 &            12.94 &             9.12 \\
\hline
\end{tabular}
\end{table}

\begin{table}
\centering
\contcaption{}
\begin{tabular}{rrrr}
\hline
    TOI &  20s Improv. [\%] &  120s Improv [\%] &  600s Improv [\%] \\
\hline
3720.01 &            13.04 &            12.94 &            11.14 \\
2026.01 &            12.64 &            12.31 &             7.96 \\
5668.01 &            12.44 &            12.23 &             9.57 \\
3877.01 &            11.89 &            11.80 &            10.40 \\
5760.01 &            11.80 &            11.59 &             8.33 \\
5364.01 &            11.38 &            11.29 &             9.77 \\
2791.01 &            11.26 &            11.15 &             9.21 \\
5594.01 &            11.10 &            10.89 &             8.06 \\
5636.01 &            11.02 &            10.04 &             2.83 \\
3694.01 &            10.89 &            10.32 &             6.28 \\
2987.01 &            10.73 &            10.14 &             6.90 \\
5590.01 &            10.56 &             9.38 &             4.91 \\
5649.01 &            10.53 &            10.47 &             9.33 \\
3683.01 &            10.47 &            10.26 &             7.73 \\
5502.01 &            10.34 &             9.92 &             5.29 \\
3761.01 &            10.25 &            10.21 &             9.30 \\
5690.01 &            10.07 &            10.01 &             8.66 \\
5595.01 &            10.00 &             9.47 &             4.33 \\
3696.01 &             9.51 &             9.28 &             6.47 \\
5755.01 &             9.34 &             9.07 &             5.93 \\
3981.01 &             9.25 &             9.14 &             7.22 \\
1560.01 &             9.25 &             9.15 &             7.79 \\
3988.01 &             9.23 &             8.98 &             5.81 \\
1546.01 &             8.58 &             7.25 &             4.46 \\
3668.01 &             8.13 &             8.04 &             6.79 \\
5645.01 &             8.05 &             7.75 &             5.26 \\
5625.01 &             7.84 &             7.04 &             1.84 \\
5113.01 &             7.40 &             7.21 &             5.36 \\
2985.01 &             7.33 &             7.24 &             5.99 \\
3717.01 &             7.33 &             7.28 &             6.66 \\
5580.01 &             7.00 &             6.88 &             5.61 \\
5582.01 &             6.27 &             6.18 &             4.91 \\
5761.01 &             6.10 &             5.99 &             4.48 \\
5776.01 &             5.96 &             5.82 &             4.09 \\
3644.01 &             5.75 &             5.71 &             4.82 \\
1593.01 &             5.27 &             5.22 &             4.33 \\
4001.01 &             5.16 &             5.02 &             3.36 \\
3244.01 &             4.76 &             4.73 &             4.21 \\
5664.01 &             3.41 &             3.35 &             2.61 \\
5599.01 &             3.00 &             2.90 &             1.76 \\
5699.01 &             2.99 &             2.96 &             2.52 \\
5637.01 &             2.83 &             2.80 &             2.17 \\
\hline
\end{tabular}
\end{table}


\bsp	
\label{lastpage}
\end{document}